# Saturn satellites as seen by Cassini Mission

**A. Coradini (1), F. Capaccioni (2), P. Cerroni(2), G. Filacchione(2), G. Magni,(2) R. Orosei(1), F. Tosi(1) and D. Turrini (1)**

(1)IFSI- Istituto di Fisica dello Spazio Interplanetario INAF Via fosso del Cavaliere 100- 00133 Roma

(2)IASF- Istituto di Astrofisica Spaziale e Fisica Cosmica INAF Via fosso del Cavaliere 100- 00133 Roma

# Table of content

## Abstract

In this paper we will summarize some of the most important results of the Cassini mission concerning the satellites of Saturn.

The Cassini Mission was launched in October 1997 on a Titan IV-Centaur rocket from Cape Canaveral. Cassini mission was always at risk of cancelation during its development but was saved many times thanks to the great international involvement. The Cassini mission is in fact a NASA-ESA-ASI project. The main effort was made by NASA, which provided the launch facilities, the integration and several instruments; ESA provided the Huygens probe while ASI some of the key elements of the mission such as the high-gain antenna, most of the radio system and important instruments of the Orbiter, such as the Cassini Radar and the visual channel of the VIMS experiment. ASI contributed also to the development of HASI experiment on Huygens probe. The Cassini mission was the first case in which the Italian planetology community was directly involved, developing state of the art hardware for a NASA mission.

Given the long duration of the mission, the complexity of the payload onboard the Cassini Orbiter and the amount of data gathered on the satellites of Saturn, it would be impossible to describe all the new discoveries made, therefore we will describe only some selected, paramount examples showing how Cassini's data confirmed and extended ground-based observations. In particular we will describe the achievements obtained for the satellites Phoebe, Enceladus and Titan. We will also put these examples in the perspective of the overall evolution of the system, stressing out why the selected satellites are representative of the overall evolution of the Saturn system.

Cassini is also an example of how powerful could be the coordination between ground-based and space observations. In fact coordinated ground-based observations of Titan were performed at the time of Huygens atmospheric probe mission at Titan on 14 January 2005, connecting the in situ observations by the probe with the general view provided by ground-based measurements. Different telescopes operating at different wavelengths were used, including radio telescopes ( up to 17- tracking of the Huygens signal at 2040 MHz), eight large optical observatories studying the atmosphere and surface of Titan, and high-resolution infrared spectroscopy used to observe radiation emitted during the Huygens Probe entry (Witasse et al. 2006).

## Introduction

The Cassini Orbiter is one of the most complex spacecraft ever developed, so it is worthwhile to describe it before attempting to summarize data analysis and results. In what follows we will describe the S/C and the main characteristics of the data sets that have been generated (section 1). A further step is the description of how these very different data sets can be combined to achieve a better understanding of the evolution of the satellites. This will be the subject of the last section of this paper.

In the second section we will describe the Saturnian satellite system and how a synthetic view of the system and of the complex interactions existing among the satellites can be achieved by its systematic study Global data obtained by Cassini can be compared with the possible overall evolution of the satellite system, as deduced by the current theories on the satellite formation and with the results of ground-based observations, and can bridge the gap between these two approaches.

In section 3 some of the most important achievements obtained for Phoebe are described: the Cassini mission has clearly revealed an "alien" chemical composition for this representative of the captured satellites, as confirmed by a comparison of Phoebe's mineralogical composition to existing spectra of KBOs.

Section 4 will be devoted to Enceladus, showing how the results from different Cassini Orbiter instruments were combined to achieve a completely new view of this satellite. The case of Enceladus is paradigmatic of how much can be achieved by the detailed analysis done by a space probe as opposed to ground based observations. Cassini with its long duration is obviously a special example, since it has been possible to combine high spatial and spectral resolution – typical of space experiments - with frequent and non- sporadic monitoring of the satellite. This can be achieved only by large missions that have to be considered the highlights of most important international Agencies; smaller and less expensive mission can monitor a planet only for a limited amount of time: in these cases, the combination of "in situ" observations with systematic ground-based observations becomes essential.

In section 5 we will describe the recent observations of Titan made mainly by the Cassini Radar and by the Visual IR Mapping Spectrometer of Cassini.

We will conclude comparing our existing knowledge on Saturn's satellites with the predictions made by formation theories and discussing how the observed features can be used to confirm or disprove the existing schemes.

## The Cassini Mission payload and data

Cassini-Huygens is a joint NASA/ESA/ASI mission to explore Saturn system, and in particular Titan and the icy moons surrounding the planet, as well as the Saturnian rings. The mission consists of two elements: the Cassini Orbiter (described here in some detail) and the Huygens probe that has been deployed into the Titan atmosphere.

The Cassini Orbiter was built by NASA's Jet Propulsion Laboratory. The Italian Space Agency contributed its high-gain antenna, the K-band subsystem and different instruments, the Visual Channel of the Visual-IR mapping Spectrometer (hereafter named VIMS) and the Cassini Radar. These contributions were regulated by a bilateral NASA/ASI agreement.

Huygens, and the associated communications equipment on the Orbiter, was built by ESA. ASI contributed to the Huygens probe via the usual ESA mandatory program contribution, while it contributed directly to the H-ASI instrument, a suite of different sensors capable of exploring Titan's atmosphere. The complex payload of the probe contained several instruments provided by the different European nations as well as by US scientists. The integration between scientists of different nationalities and background during the Cassini mission contributed in a major way to the success of the mission.

The Cassini-Huygens spacecraft was launched on October 15$^{th}$, 1997 by a Titan IVB-Centaur rocket from Cape Canaveral. The Spacecraft wasn't injected into a direct trajectory to Saturn due to its weight and complexity, but made use of gravity-assisted manoeuvres at Venus, Earth and Jupiter. These manoeuvres increased the duration of the voyage, which lasted about 6.7 years, but allowed to test the instruments during the different fly-bys and to improve their calibration. As an example, the VIMS calibration was improved by using the Venus data and lunar data.

On July 1$^{st}$, 2004 the Cassini-Huygens spacecraft entered the Saturnian system starting the nominal mission, that lasted four years and that is now concluded. On December 25th, 2004, the Cassini Orbiter released the Huygens probe, which started its 20-days long trip to Titan. The probe started its observations on January 14$^{th}$, 2005. The descent phase was regulated by a parachute and lasted around 2 hours and half. The probe was not destroyed when impacting the surface, which was solid and not liquid as expected, and survived for about an extra hour. The first four years of the Cassini-Huygens measurements were fundamental to understand the complexity and the diversity of Saturn's system. At the end of the Nominal mission a Cassini Equinox Mission started, which will last for the following two years; the name stresses out the fact that Cassini Orbiter will be able to observe seasonal changes in the atmosphere at Saturn and Titan.

Unique ring events will be observed as well, such as waves in the rings, during the 2009 equinox passage of the Sun through the plane of the rings. During these two years the spacecraft will make 60 additional orbits of Saturn, including 26 flybys of Titan, seven of Enceladus, and one each of Dione, Rhea and Helene.

In table 1 the Cassini Orbiter Payload is summarized: we will briefly describe it and the kind of information that is essential for our discussion. We will not describe the Huygens probe since we are interested here in the system of satellites as a whole and not in the specific in situ investigation of Titan's atmosphere. The Cassini instruments are grouped into suites.

## The Cassini payload

The optical instruments are mounted on the remote sensing pallet and are therefore aligned: their goal is to study Saturn and its rings and moons in the electromagnetic spectrum. These instruments are those whose data can be better compared with those from ground-based observations: the Composite Infrared Spectrometer (CIRS), the Imaging Science Subsystem (ISS), the Ultraviolet Imaging Spectrograph (UVIS) and the Visible and Infrared Mapping Spectrometer (VIMS). The microwave instruments, using radio wavelength, have been included to map atmospheres and the surface of Titan despite its dense atmosphere. Radio Science experiments are used to determine the mass of moons and collect data on the size distribution of ring particles.

The Fields, Particles and Waves instruments study the dust, plasma and magnetic fields around Saturn. While most don't produce actual "pictures," the information they collect is critical to understand the Saturnian environment. They are considered "in situ" instruments.

The Microwave Remote Sensing, Using radio waves, these instruments map atmospheres, determine the mass of moons, collect data on the size distribution of ring particles and unveil the surface of Titan. They are the Radar and Radio Science (RSS)

Here we will mainly discuss the remote sensing instruments data and will show how they can be used to infer both general characteristics of the satellites and specific aspects of Saturn's satellites. In particular, given our experience in imaging spectroscopy, we will discuss with greater care the VIMS data, which proved to be extremely useful in understanding the relationships between different satellites. The Cassini data are in PDS format. The Planetary Data System (PDS) is an archive of data products for NASA and ESA planetary missions. All PDS products are peer-reviewed, documented, and easily accessible via a system of online catalogues that are organized by planetary disciplines.

## Satellites origin and bulk characteristics

The study of the formation conditions of the Saturnian icy satellites can be the key to understand the formation of Saturn itself and of its surrounding disk. Irregular satellites are instead believed to have formed in the solar nebula and to have been gravitationally captured by the giant planet at a later time. Irregular satellites can supply information on the conditions of the surrounding space, as well as on the characteristics of a planet that was able to trap bodies with a higher efficiency than the present Saturn. Connecting a model for the solar nebula to a model of giant planet formation including the formation of regular satellite systems is quite a difficult problem. Using a 3-D hydro-dynamical model, Magni and Coradini (2004) showed that, in the last stage of giant planet formation, a proto-satellite disk of captured material gradually emerges from the contracting atmosphere. Satellite formation could have taken place in such a disk, whose evolution is controlled by the last phases of the giant planet formation process (Alibert, Mousis and Benz, 2005). Moreover, the amount of material available to form the regular satellites depends closely on the amount of gas and solids accreted and/or captured by this sub-nebula.

The sub-nebula is fed –at the beginning of the process of planetary formation- by gas particles originating from the solar nebula, penetrating in the nebula mainly through its outer edge. During this period, the temperature and pressure conditions of the Saturnian sub-nebula reach their maximum value. When large part of the gas content of the solar nebula has been dissipated, the mass flux toward the sub-nebula stops and the Saturnian sub-nebula gradually empties. A large fraction of the gas is either dissipated or accreted by the forming Saturn. At the same time, due to angular momentum conservation, the sub-nebula expands outward. Such an evolution implies a rapid decrease of temperature, pressure and surface density over several orders of magnitude in the whole Saturnian sub-nebula. An example of the density distribution in the Saturn sub-nebula is given in Figure 1**.** The sub-nebula passes through different stages of turbulence, which can be modelled by assuming different values of the disk viscosity that is described by the so-called α parameter (Alibert and Mousis, 2006). Coradini et al. (2009) have followed the disk evolution in different phases by using this strategy. The initial phases are characterized by high values of disk viscosity, and as soon as the disk evolves, the viscosity decreases and the sub-nebula gradually becomes quiescent. Figure 2 shows the disk phase diagram at different times: such figure can be considered as representative of the evolution diagram of the disk. Under the thermodynamic conditions predicted by these models, some ices/rocks gradient in regular satellites with increasing distance from Saturn should be observed, if satellites formed in the Saturnian sub-nebula at early epochs or formed in a solar nebula with a high volatile content. According to Coradini et al. (2009) models, water ice as well as ammonia ice is always condensed in the Saturnian disk while carbon monoxide is condensed only at the end of Saturn's accretion (see Figure 3). Carbon monoxide, however, could be further incorporated, once the gas of the disk is in large part dissipated, through the planetesimals injected in the disk. The chemistry of the disk evolves in time and the amount of condensed material with respect to the gas evolves as well.This scenario predicts in particular that some volatile species should have been incorporated in forming satellites (Mousis & Alibert 2006).

Spectral observations of Saturn from the far infrared spectrometer onboard the Cassini spacecraft (Flasar, F.M., et al., 2005) revealed that the C/H ratio in the planet is enriched by a factor of almost 11 (Fletcher et al. 2009) with respect to solar composition (Grevesse et al., 2007). Moreover, this enrichment is almost twice than that of Jupiter. These observations are an indication of the fact that the gas incorporated by the giant planets was already partially depleted in hydrogen, thus indicating a later accretion of Saturn with respect to Jupiter. This difference in the formation time and therefore in the *C* content of the two forming planets reflects the difference in the accretion disks from which the satellites formed Part of the carbon could have been incorporated by the satellites at the time of their formation, but the part in the form of CO ice could have been added also at a later time from external sources. Carbon-rich materials could be incorporated yet and still see as contaminants. Filacchione et al. (2006) performed a comparative analysis of more than 1400 full-disk observations obtained from July 2004 to present by Cassini for 15 between regular and minor satellites. These observations, carried out from the equatorial plane, are particularly suitable to highlight the spectral differences among the different satellites and to shed light on systematic effects. The combined use of several VIS and IR spectral quantities (e.g. spectral slopes, water ice bands strengths, continuum levels, etc.) allows finding correlations between classes of satellites orbiting at different distances from Saturn. In this way it has been possible to discriminate the almost pure ice surfaces of Enceladus and Calypso from the organic rich Hyperion, Iapetus and Phoebe (Filacchione et al. 2008).

Water ice is abundant on Saturn's satellites, and near-infrared spectra returned by VIMS can be used to study the crystalline order of the surface ice on all the satellites. It could also be a way to measure the lattice order of ice that –in turn- depends on its condensation temperature and rate, its temperature history, and its radiation environment. Inner satellites are characterized by a larger content of crystalline ice, with respect to the amorphous one (which is almost absent) -measured on the basis of the strength of the absorption band at 2 and 2.05 μm (see spectra collected by Cassini VIMS.

Figure s 4, 5 and 6). Enceladus is an exception due to the fact that new "fresh" ice is generated on its surface. Even if the content of amorphous/crystalline ice isn't an unambiguous parameter, it gives a clear indication of the temperature gradients existing in the solar nebula since amorphous ice should have formed at lower temperature then crystalline ice.

A similar analysis performed on CO2 ice yields the radial distribution of the "contaminants" across the Saturnian system, measured through the visible spectral slopes (angular coefficient of the best fit line to the spectrum normalized at 0.55 μm) and on the strength of the 4.26 micrometers absorption band. These results are obtained from a statistical analysis of more than 1400 full-disk observations of the Saturnian icy satellites obtained by Cassini-VIMS during the nominal mission (Filacchione et al., Icarus 186, 2007; Filacchione et al., Icarus, submitted).

The satellites have red (positive) slopes in this range, probably caused by the presence of a still unidentified UV absorber. In the inner region the minimum is observed on the more uncontaminated object (Enceladus) reaching a maximum on Rhea and apparently decreasing from Hyperion to Phoebe. The radial distribution of the spectral slope is sensitive to both regolith grain size and degree of contamination: in this case we observe negative values on Enceladus (similar to pure water ice surface), neutral on Mimas, Tethys and Dione, maximum on Hyperion and strongly positive on Iapetus (Filacchione et al. 2007) **(Figure 7).** From this analysis, it is clear that satellites underwent extensive contamination from external agents, rich in carbon compounds.
Only thanks to the Cassini mission it has been possible to acquire a global view of the satellites as a whole.

## Selected bodies

### Phoebe

Phoebe is the ninth satellite of Saturn and it is characterized by several peculiar properties. Its retrograde orbit, together with its great semimajor axis (12.952 million km or 214.9 Saturn radii, more than10 times that of Titan) and its relatively high eccentricity (e = 0.1644) and inclination (i = 174.75°), suggest that this moon was captured (Pollack et al., 1979) and thus “irregular”. Seen from the Earth, Phoebe’s spectrum is essentially flat and grey in the visual region (Tholen and Zellner, 1983; Buratti et al., 2002); its albedo is higher than that due to carbon, particularly in the brightest areas detected by Voyager 2. In the infrared, instead, Earth-based spectroscopic observations showed the typical signatures of water ice (Owen et al., 1999; Brown, 2000) suggesting a relationship with the icy bodies of the outer Solar System, such as Kuiper Belt objects, Centaurs or comets. All these data were somehow difficult to interpret since the former suggest a chondritic composition for Phoebe while the latter an icy one. These ambiguities were solved during the flyby of Cassini-Huygens (11 June 2004), whose distance of closest approach was 2068 km from Phoebe. To that purpose, the role of the camera and that of VIMS, which collected data with spatial resolution at best of ~ 1 km/pixel, were fundamental (see table 1). In the images acquired by the Cassini multispectral camera, Phoebe is characterized by a heavily cratered surface, with overlapping craters of varying sizes. This morphology suggests an old surface. There are apparently many craters smaller than 1 km, indicating that projectiles probably smaller than 100 meters once bombarded Phoebe. All images show evidence for an ice-rich body coated with a thin layer of dark material. The VIMS spectrometer collected several cubes, unfortunately with different spatial resolutions, phase angles and integration times (Coradini et al.

2007). A first analysis of Phoebe mineralogy, deduced from the VIMS data, was performed by Clark et al. (2005) and showed clearly the complex composition of Phoebe and the presence of "organic ices". The spectra of Phoebe indicate a low albedo surface, from 1% to 6% reflectance **(Figure 8)**, with a variety of absorption features probably due to materials occurring with variable abundances and/or grain sizes at different locations on the body. They include: previously identified water ice, bound water, and trapped CO2 (**Figure 8**). A broad 1-mm feature, attributed to Fe2þ-bearing minerals, is present in almost all regions but is stronger in equatorial areas (Clark et al. 2005). A more complete analysis of Phoebe's mineralogy using multivariate classification of areas similar from a mineralogical point of view has shown that complex hydrocarbons, identified by the presence of bands at 3.5 and 4.4 μm (Coradini et al. 2008), are possibly present on Phoebe, thus supporting the hypothesis that Phoebe is a captured object.

## Enceladus

Enceladus, only 504 km in diameter, is a complex world from a geological point of view. Earlier Earth-based and Voyager space probe images revealed that Enceladus' surface is icy and complex. Old cratered terrains are interlaced with newly resurfaced smooth ice flows. Kargel (2006) stresses out that Enceladus once was believed to be too small to be active. Thanks to Cassini data, it has been found to be one of the most geologically dynamic objects in the Solar System. From a compositional point of view Enceladus' surface is composed mostly of nearly pure water ice except near its south pole, where light organics, CO2, and amorphous and crystalline water ice have been observed. CO is also present in the atmospheric column above Enceladus (Brown et al. 2006). The spectrum of Enceladus is bright and featureless, like pure water ice, and the near-IR spectrum has also been compared to pure water ice (Hendrix and Hansen, 2009, and references therein) or pure water ice plus a small amount of NH3 hydrate or NH3 (ibid). FUV measurements of dark regions on Enceladus can be explained by the presence of a small amount of NH3 and a small amount of a tholin in addition to H2O ice (Hendrix and Hansen, 2009).Cassini identified also a geologically active region at the south pole of Saturn's moon Enceladus. In the images acquired by the Imaging Science Subsystem (ISS) (Porco et al. 2006) this region is surrounded by a chain of folded ridges (named "tiger stripes"). Southward of this boundary a region of very different albedo and high temperatures exists. The folded ridges region is also characterized by the presence of crystalline ice, revealed by VIMS (Brown et al. 2006). Cassini's Composite Infrared Spectrometer (CIRS) detected 3 to 7 gigawatts of thermal emission from the same south polar troughs at temperatures up to 145 Kelvin or higher (Spenser et al. 2006). New more precise measurements of CIRS revealed temperatures of at least 180°K, which is 93°K warmer than other regions of the moon, and 17°K higher than previously measured on the same location. The already warm region of the tiger stripes is characterized by the presence of two "hot spots" in correspondence to two of the jets.

Enceladus –again from the same region- is emitting jets of organic-laden water vapour and dust-sized icy particles from these regions. These plumes supply material to Saturn's E-ring, "replenishing" it. During Cassini's close flyby of Enceladus on July 14th, 2005, the High Rate Detector of the Cosmic Dust Analyzer registered micron-sized dust particles enveloping this satellite. The dust impact rate peaked about 1 minute before the closest approach of the spacecraft to the moon. This asymmetric signature is consistent with a locally enhanced dust production in the south polar region of Enceladus. Therefore, thanks to all these measures performed by different instruments, it is clear that the source of E-ring's dust and gas was Enceladus (Sphan et al. 2006). INMS -the Cassini Ion and Neutral Mass Spectrometer- during two close flybys of Enceladus revealed the presence of ammonia, complex organics such as benzene, and deuterium in the gas plume as well as the probable presence of radiogenic argon (Waite et al, 2009). According to these authors, there is strong evidence, in the INMS data, of liquid water in Enceladus' interior. Moreover the measurement of the ratio of deuterium to hydrogen seems to be consistent with the accretion of Enceladus from planetesimals formed in the solar nebula in the region of the giant planets, as suggested by Coradini et al, (2009).

It has been noted that the Cassini magnetometer detected the interaction of the magnetospheric plasma of Saturn with an atmospheric plume. This discovery was made during a distant flyby and was later confirmed during two follow-on flybys. The magnetometer data are consistent with local out-gassing activity via a plume from the surface of the moon near its south pole, as confirmed by other Cassini instruments (Dougherthy et al., 2005). Therefore the Enceladus plumes not only replenish the E-ring, but are also responsible for locally modelling the Saturnian magnetosphere.

The investigation of Enceladus is a typical example of how all the instruments have to be used in order to achieve a new understanding of the evolution of the icy moons, since each of them represents a new world by itself and a new challenge. However, little is known today about the mechanisms generating the intense activity of Enceladus south pole. Nimmo and Pappalardo (2006) suggested that this heat flux is probably due to localized tidal dissipation within either the ice shell or the underlying silicate core. The surface deformation is possibly due to upwelling of low-density material (diapirism) as a result of this tidal heating. However the problem is not yet totally solved. Roberts et al (2009) accurately modelled Enceladus thermal evolution, which seems to be not incompatible with the presence of an ocean, yet indicate that this ocean should have formed under different geophysical conditions than the present ones and in the presence of a certain amount of ammonia ice to lower the freezing point of the mixture ( Roberts et al. 2009).

## Titan

Before the Cassini-Huygens spacecraft arrived at the Saturnian system, very little was known about Saturn's largest moon Titan. Ground-based observations and Voyager data had revealed a thick atmosphere composed primarily of nitrogen with a small percentage of methane and higher order hydrocarbons. Titan is the only satellite in the Solar System to have a dense atmosphere, composed primarily of nitrogen with a few percent of methane. This atmosphere is completely opaque at visible wavelengths, due to the absorption of aerosols and gas and the scattering by aerosols.  The combined observations made by Cassini and by the largest telescopes and radio-telescopes in the word made at the time of Huygens deployment allowed to measure temperature profiles between 400 and 600 km height (Witasse et al. 2006). Thanks to these measurements it was identified the sharp inversion layer near the 515 ± 5 km altitude level. At that level, the temperature locally increases by 15 K in only 6 km, and the peak value of the gradient dT/dz reaches values as high as +6 K/km. This layer has also been observed by the HASI experiment aboard Huygens, at around 507 ± 15 km. Further work is needed to understand the difference in altitude. In general a new knowledge of the Titan atmosphere was achieved, thanks to these combined data, permitting also to improve the so-called "engineering model" of the Titan atmosphere. Vertical distribution of ethane was also measured (ibid).  NIR images from Keck showed differences between high reflectance areas and potential areas covered by hydrocarbons (Witasse et al, 2006 and references therein). However, given to the complexity of the Titan atmosphere and the amount of aerosols, from ground based observations, the access to the Titan surface was very limited.  In fact the surface is obscured by hydrocarbon smog at the visible wavelengths, thus the Titan Radar Mapper was included in the scientific payload of the Cassini spacecraft to study the surface of Titan. Indeed, Titan's dense atmosphere has been penetrated by Cassini-Huygens to reveal a complex world with diverse geophysical and atmospheric processes reminiscent of those on Earth, but operating under very different conditions from our home world. Titan also possesses a methane cycle remarkably analogous to the hydrological cycle on the Earth but with the fluid oceans removed. Fluvial features, lakes, clouds and tentatively rain may exist, but the working fluid is methane and its photochemical product ethane. Tantalizing but circumstantial evidence exists for a form of volcanism involving water ice and possibly ammonia. In the upper atmosphere (up to at least ~1100 km), photochemistry driven by ultraviolet light and the chemistry of the charged particles produce suites of complex heavy hydrocarbons and nitriles, affecting the thermal balance and chemistry of the whole atmosphere. Cassini data suggest that Titan's climate has not been static over much of the history of the Solar System and may still be changing today. The ultimate source of methane and the sink for its organic products are key outstanding questions about this cryptic world.

In exploring the surface of Titan, an extremely powerful combination of data from the Cassini Orbiter instruments is the joint coverage by the multi-mode RADAR Investigation and the Visible and Infrared Mapping Spectrometer (VIMS). The Cassini Titan RADAR Mapper operates at Ku-band (13.78 GHz frequency or 2.17 cm wavelength) and collects low-resolution (several to tens of km) scatterometer, altimeter, and radiometer data as well as very high-resolution (down to ~350 m) synthetic aperture radar (SAR) images covering large strips of Titan's surface (Elachi et al., 2004). VIMS collects spectral cubes that are more limited in spatial coverage, at best usually a few km in resolution (and rarely a few hundred meters in resolution), but covers a large spectral range from 0.35 to 5.2 μm (Brown et al., 2004). VIMS provides information about the morphology of the surface thanks to seven infrared windows, where the atmospheric methane is not effective as an absorber. In particular, very sharp images are acquired at 2 μm, where the scattering by aerosols is much reduced compared to the shorter wavelengths. Finally, the ISS camera (Porco et al., 2005) provides images of the surface in the 0.93 μm filter, once the strong scattering by aerosols has been corrected by image enhancement techniques (Perry et al., 2005). Cassini has discovered features such as seas of dunes, lakes in the polar regions, and a young surface marked by few craters. Images from the Titan Radar Mapper in its Synthetic Aperture Radar (SAR) mode, together with those from the Imaging Science Subsystem (ISS) and the Visual and Infrared Mapping Spectrometer (VIMS), revealed a complex surface that has been shaped by all major planetary geologic processes - volcanism, tectonism, impact cratering and erosional/depositional processes, both from fluvial and aeolian activity.

## Dunes

Fields of dunes (Titan's "sand seas") are mostly equatorial, but a few isolated patches of dunes extend as far north as ~60 degrees (Lorenz et al., 2006). They are thought to be composed of small hydrocarbon or water ice particles - probably about 250 microns in diameter, similar to sand grains on Earth. These are formed into dunes by the prevailing west-to-east surface winds. Because of the shape and length of the dunes, they are probably longitudinal rather than transverse dunes, which form across the wind and are more common on Earth.

## Craters

More than 30% of the surface of Titan has been imaged by Cassini Radar through data taken in T44, but seven surface features have been unambiguously identified as impact structures (Lorenz et al., 2007, Wood et al., 2009), while fifty additional craters have been tentatively identified. Compared with Saturn's other moons with their thousands of craters, Titan's surface is very sparsely cratered. This is in part due to Titan's dense atmosphere, which burns up the smaller impactors before they hit the surface, but also to Titan's active geologic processes, such as wind-driven motion of sand and icy volcanism, erasing craters over time. Craters fall into two distinct morphologies. The most common is similar to that of craters from the Moon and Mars, with an extensive ejecta blanket, a raised rim, and a large central peak. Other craters have radar-bright and apparently jagged rims with smooth floors, and lack terraces and other indications of collapse. This hints at the possibility that crustal properties or thicknesses vary across Titan (Wood et al., 2009).

## Fluvial features

Features resembling terrestrial rivers have been seen in SAR images since the first close flyby of Titan, and have been found to be common on its surface. They are often several hundreds of kilometers in length, and their morphology and backscattering characteristics are correlated with latitude. Fluvial features at low- and mid-latitude are often braided and brighter than the surrounding terrain. Only a few radar scattering mechanisms can explain such high radar returns. The presence of rounded, radar-transparent (icy) pebbles with size larger than the radar wavelength could result in an enhanced radar cross section: these radar-bright channels are thus interpreted as river beds where debris, likely shaped and transported by fluvial activity, have been deposited (Le Gall et al., in preparation). Most fluvial features at high latitudes are radar-dark and meandering, and connect polar lakes. Their low radar backscatter cross section can be explained by the presence of liquid hydrocarbons or fine-grained sediments, similarly to lacustrine features. A third type, seen predominantly at mid- and high latitudes, has a characteristic radar brightness profile, with a bright downrange bump paired with a dark uprange trough, denoting appreciable negative relief. These features are up to 3 km across, while channel wall slopes estimated from radarclinometry imply a depth of incision of several hundred meters (Lorenz et al., 2008a).

## Tectonic features

Cassini has observed both isolated blocks and linear chains of mountains 200 - 300 km long on Titan’s surface. Maximum elevations of about 2000 m have been derived from radar images through a shape-from-shading model corrected for the probable effects of image resolution (Radebaugh et al., 2007). Erosion rate estimates for Titan provide a typical mountain age as young as 20-100 million years (Radebaugh et al., 2007). Possible formation mechanisms for mountains include crustal compressional tectonism and upthrusting of blocks, extensional tectonism and formation of horst-and-graben, deposition as blocks of impact ejecta, or dissection and erosion of a pre-existing layer of material. All above processes may be at work, given the diversity of geology evident across Titan's surface. Modeling of the formation of mountains through compressional crustal deformation, as a consequence of Titan’s radial contraction during hot periods of its thermal evolution, produces a topography height of several kilometers (Mitri et al., 2008).

## Lakes

For more than two decades, scientists have debated whether liquids on Titan exist, and if so, where they would be located. Pre-Cassini observations from the 1980s indicated that something on Titan's surface must be re-supplying the methane observed in its atmosphere, which is continuously destroyed by ultraviolet radiation, and a global ocean was once hypothesized. Subsequently, disconnected lakes or seas were predicted. The discovery of numerous lakes near Titan's north pole by the Cassini radar instrument has confirmed the latter idea (Stofan et al., 2007), but it has not solved the puzzle of the origin of the methane, which the lakes are a sink for because the atmospheric circulation on average transports material from equator to pole. Many apparently empty lake basins are seen in radar data, especially at high southern latitudes, while there seems to be a preference during the current season for liquids to be located near the north pole (Fig. 1). Since Titan is currently in its late summer season in the southern hemisphere, this interpretation is consistent with a previously proposed theory that methane fills the lakes during the winter and evaporates during the summer, leaving them dry until the next fall.

## Cryovolcanism

It has been long suspected that Titan may have cryovolcanoes, and the Cassini mission has collected data from several flybys of the moon that suggest their existence (Lopes et al., 2007). Imagery of the moon included a suspect haze hovering over flow-like surface formations that some point to as signs of possible cryovolcanism. Cryovolcanic units have been identified in SAR images mostly at mid-latitudes (40°-60° N), these include the construct Ganesa Macula, several calderas with associated flows, and large cryovolcanic flows. Some features, identified as cryovolcanic lava flows in SAR images, have also a variable infrared brightness in VIMS images, suggesting that these are recent cryovolcanoes (Wall et al., 2009). The composition of the cryomagma on Titan is still unknown, but constraints on rheological properties can be estimated using flow thickness. Rheological properties of one flow were estimated and appear inconsistent with ammonia water slurries, and possibly more consistent with ammonia water methanol slurries (Lopes et al., 2007).

## Asynchronous rotation

Cassini radar observations of Saturn's moon Titan over several years show that its rotational period is changing and is different from its orbital period. The present-day rotation period difference from synchronous spin leads to a shift of ~0.36° per year in apparent longitude and is consistent with seasonal exchange of angular momentum between the surface and Titan's dense super-rotating atmosphere. This phenomenon had been predicted by Tokano and Neubauer (2005), and its extent can be explained only if Titan's crust is decoupled from the core by an internal water ocean like that on Europa (Lorenz et al., 2008b).

After the end of the primary mission, Cassini's two-year "Equinox" extended mission has begun, and there are hopes for a possible six-year Cassini "Solstice" Mission to follow. Cassini/Huygens arrived at Titan at the equivalent of Earth's January and it will be June on the satellite by the time the Solstice Mission will be completed. Fundamental questions that will be addressed by the extended mission include the determination of the presence of a liquid layer in Titan's interior. Gravity passes and radar swaths are needed to answer this question. To date no interior magnetic field has been detected, and a very close Titan flyby is planned at the end of the Equinox mission to go below the ionosphere. Titan's atmosphere has weather that can be compared to that on Earth, with the formation of clouds and rainfall. The over-arching goal for the Cassini Solstice mission would be to observe changes, both seasonal and due to any other surface activity. The Cassini/Huygens mission is slowly revealing a complex, new world. Infrared spectroscopic data, obtained by the Visual and Infrared Mapping Spectrometer (VIMS) on board the Cassini spacecraft strongly indicate that ethane, probably in liquid solution with methane, nitrogen and other low-molecular-mass hydrocarbons, is contained within Ontario Lacus, one of the largest liquid bodies of Titan (Brown et al., 2008).

Cross comparison between ISS, VIMS and RADAR images have been undertaken in order to study several regions using as much available data as possible. Global VIMS and RADAR data comparisons have been performed in order to check for systematic correlation related to surface properties (Soderblom et al., 2007; Barnes et al., 2007a). Soderblom et al. (2007) found that RADAR-dark longitudinal dune fields, seen in equatorial to mid-latitudes SAR images, are highly correlated with VIMS "dark brown" units in RGB colour composites made from the 2.0 μm, 1.6 μm and 1.3 μm images. This dark-brown unit shows less evidence of water ice, and it is clearly one of the end members of the surface. On the other hand, water ice as one of the abundant compositional end members of Titan's surface materials is certainly reasonable; this unit is in fact represented by the "dark-blue" material, whose spectral trend is consistent with water ice and thus likely forming a substrate that is enriched in water ice relative to the rest of the surface. However, the correlation between the two VIMS and RADAR data sets is not systematic in the most general case. The absence of correlation between RADAR and VIMS bright units, found in particular on the Huygens landing site (observed by both instruments, with a 15 km/pixel resolution by VIMS), was interpreted as the result of an optically thick bright mantling which might be transparent to the radar. Though the source of this bright end member material is debated, as a plethora of solid organics are expected to form in the upper atmosphere from energetic chemistry, Soderblom et al. (2007) hypothesize that a reasonable candidate is a mantling deposit of aerosol dust that might include acetylene and other simple hydrocarbon solids, whereas the dark-brown, water ice-poor end-member of the dunes could have a higher concentration of the more complex hydrocarbons and/or nitriles.

Barnes et al. (2007b) investigated the relationships between VIMS and RADAR imagery on an equatorial region east of Xanadu using data from the ninth flyby (T9 on 26 December 2005) and eighth flyby (T8 on 28 October 2005) respectively. Channels have been observed in the two data sets, showing that VIMS was able to detect channel materials despite sub-pixel channel measured widths (~1 km). Especially near their mouths, the explored channels share spectral characteristics with Titan's dark blue terrain, consistent with an enhancement of water ice. On this point, Barnes et al. (2007b) hypothesize that the soluble portion of the organic haze that settles into the surface could be mobilized by methane rainfall and preferentially washed into channels and then out into the dark blue spectral unit, leaving behind the insoluble portion. Barnes et al. (2007b) also identified that, in the explored region east of Xanadu, areas shown to be mountainous by RADAR appear darker and bluer than surrounding terrain when observed by VIMS. They interpret this spectral variation as the result of a thin surface coating that might be present on the surrounding equatorial bright terrain but might be diminished in extent or depth, or even entirely absent within the mountainous areas.

Only a few impact craters (or circular features possibly related to impact processes) have been unambiguously detected on Titan by the Cassini-Huygens mission during its nominal mission, which indicates that the surface of Titan is geologically young (Elachi et al., 2005; Porco et al., 2005; Lorenz et al., 2007). Among these craters, Sinlap is the only one that has been observed both by the RADAR and VIMS instruments. Le Mouélic et al. (2008) have shown that interesting correlations can be observed between the spectrally distinct areas identified in the infrared data and the SAR image. Several units appear in VIMS false colour composites of band ratios in the Sinlap area, suggesting compositional heterogeneities. The dark (in infrared) crater floor corresponds to the unit delimited by the crater rim in the SAR image, with possibly a central peak identified in both. Both VIMS ratio images and dielectric constant measurements suggest the presence of a dark bluish parabolic area enriched in water ice around the main ejecta blanket. Since the Ku-band SAR may see subsurface structures at the meter scale, the difference between infrared and SAR observations can be explained by the presence of a thin layer transparent to the radar.

In general, the correlation between infrared and radar features is not systematic, and differences can be explained considering the different sensitivities of both instruments with respect to surface composition and roughness, and also the sub-surface imaging capabilities of the RADAR instrument. The two data sets provide very complementary information about surface properties, at nearly all scales. Tosi et al. (2009) applied a multivariate statistical analysis (the G-mode) to a set of data made up by infrared spectra acquired by VIMS and scatterometric echoes and brightness temperatures measured by RADAR. In such medium resolution data, sampling relatively large portions of the satellite's surface, regional geophysical units matching both the major dark and bright features seen in the optical mosaic can be pointed out. Given their data set, the largest homogeneous type is associated with the dark equatorial basins (dune fields) like Shangri-La, showing the lowest reflectance in all of the sampled atmospheric windows while having a higher brightness temperature (as expected, given the higher emissivity) and a rather low backscattering coefficient (typically $<0.15$ on average) that is most likely indicative of surfaces being relatively smooth on a regional scale. The Xanadu bright continental feature is one of the most interesting geophysical units of Titan: this region appears bright in all of the infrared atmospheric windows sampled by VIMS and it also shows the highest backscattering coefficient of the whole satellite (>0.6 on an average), consistent with a relevant roughness on a regional scale and/or with a broad volume scattering effect (Janssen et al., 2008).

## Concluding remarks

Cassini discoveries are so extensive and important that it is very difficult to summarize them in a single paper. Each of the person that contributed to this mission, were responsible for only a limited – even if valuable – contribution.

The new knowledge of the Saturn system was born by the continuous work of several scientists and technicians: several students devoted their work to the interpretation of the obtained data. It is a clear example of how an international mission should be.

The Saturn satellites are a strikingly heterogeneous ensemble: each of them, if considered as a single body, is unique. However, a systematic investigation of the Saturnian system shows that there are guidelines which allow to globally interpret its features. From a compositional point of view, Saturn's satellites are characterized by a higher content of high volatility ices than Jovian satellites. After a long debate about its presence, ammonia ice seems to be identified, at least on Enceladus. The presence of ammonia is essential in the modeling of the thermal evolution of the satellites, and can contribute to the formation of subsurface oceans (as suggested for Titan and Enceladus). Another characteristic is the presence of a strong bombardment history, which can be responsible for the lack of density gradients in this satellite system as instead found in the Jovian system.

There are also new evidences (Waite, 2009) that the satellites are formed from "local" material, i.e. present in Saturn's accretion disk or in the nearby regions. The presence of organic ices has also been revealed. Depending on the different thermodynamic conditions, this complex chemistry gives rise to new exotic material whose presence is one the main features discovered on Titan. However, not even Cassini and Huygens probe were able to completely unveil the surface composition of Titan and the peculiar characteristics of Titan's lakes, as suggested by Moriconi et al. ( 2009). This is the reason why, in the framework of ESA Cosmic Vision program, the mission Tandem, targeting Titan and Enceladus, was considered extremely valuable. In the mean time, waiting for a new mission, continuous ground-based observations can give a new insight on the atmospheric dynamics of Titan.

## Acknowledgements

This work is supported by an ASI grant.

Table 1 Cassini Orbiter Payload

| Instrument | Investigation |
|---|---|
| Imaging science subsystem | Imaging in visible, near-ultraviolet, and NIR |
| Visual and infrared mapping spectrometer | Identifies the chemical composition of the surfaces, atmospheres, and rings of Saturn |
| Composite infrared spectrometer | Measures infrared energy from the surfaces, atmospheres, and rings |
| Ultraviolet imaging spectrograph | Measures ultraviolet energy from atmospheres and rings to study their structure, chemistry, and composition. |
| Cassini radar | Maps surface of Titan |
| Radio science subsystem | Searches for gravitational waves; studies the atmosphere, rings, and gravity fields of Saturn and its moons. |
| Magnetospheric imaging instrument | Images Saturn's magnetosphere and measures interactions between the magnetosphere and the solar wind |
| Cassini plasma spectrometer | Explores plasma within and near Saturn's magnetic field. |
| Ion and neutral mass spectrometer | Examines neutral and charged particles |
| Radio and plasma wave science | Investigates plasma waves natural emissions of radio energy, and dust. |
| Cosmic dust analyzer | Studies ice and dust grains in and near the Saturn system. |

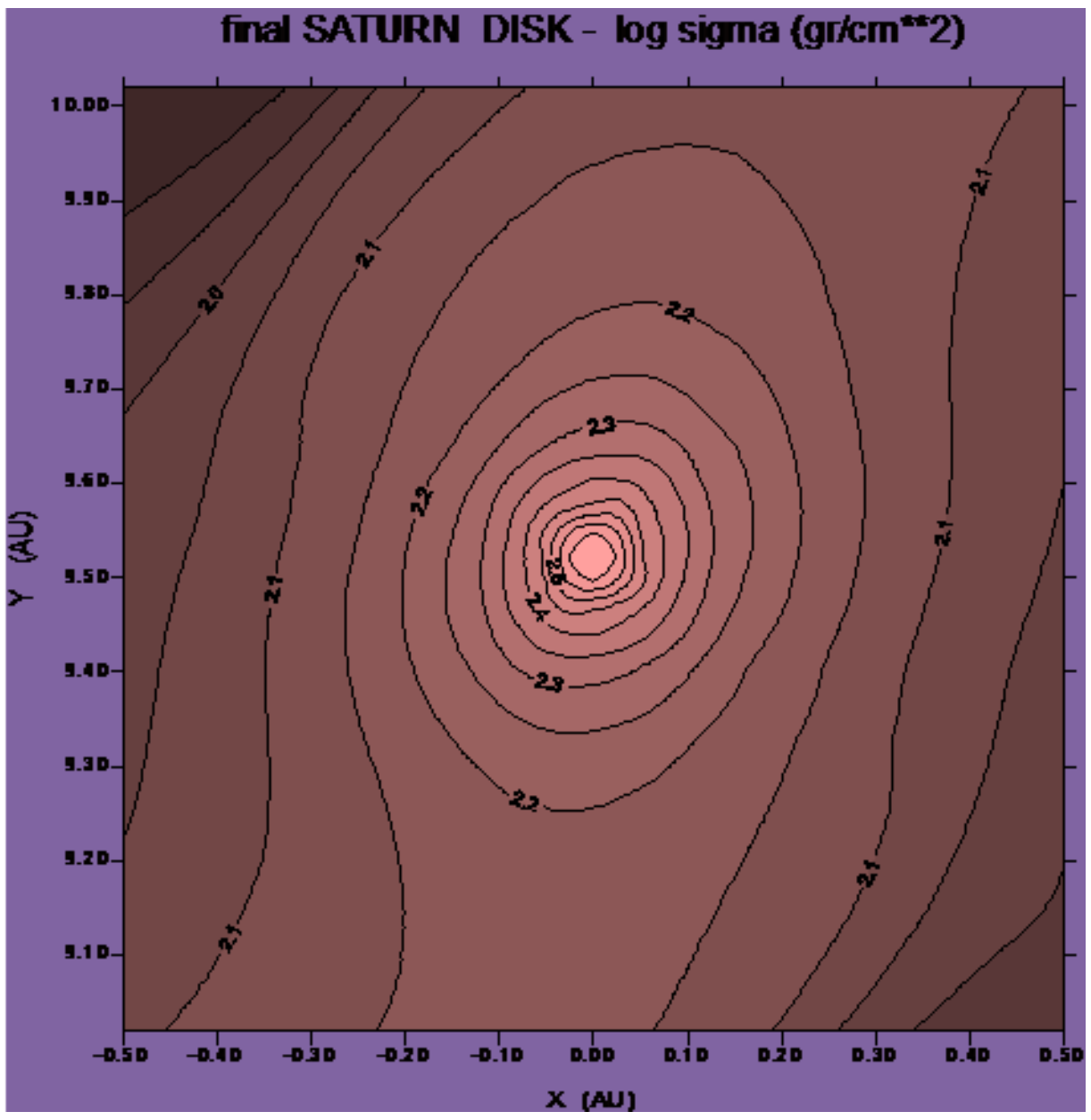

***Figure 1***

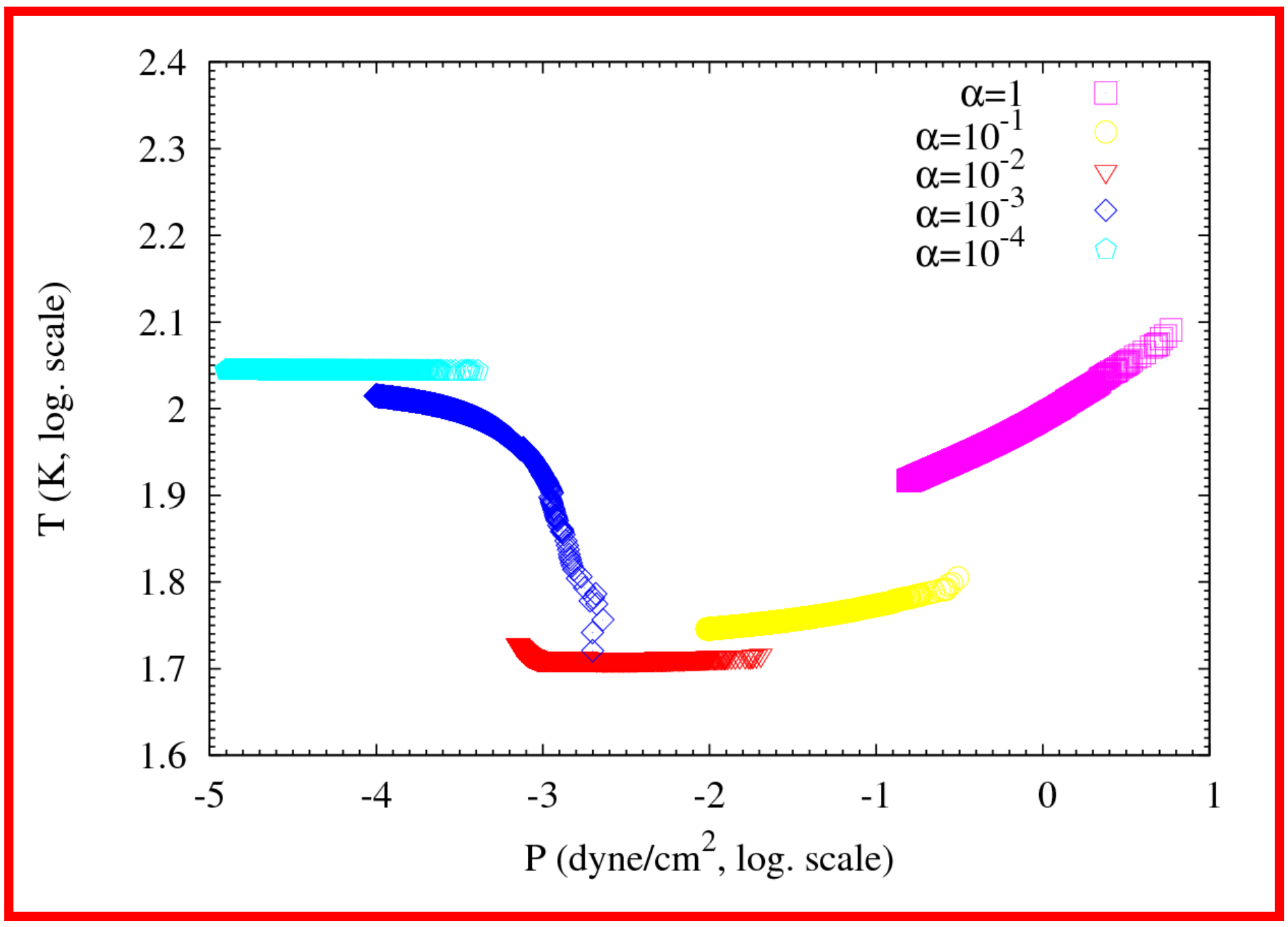

***Figure 2***

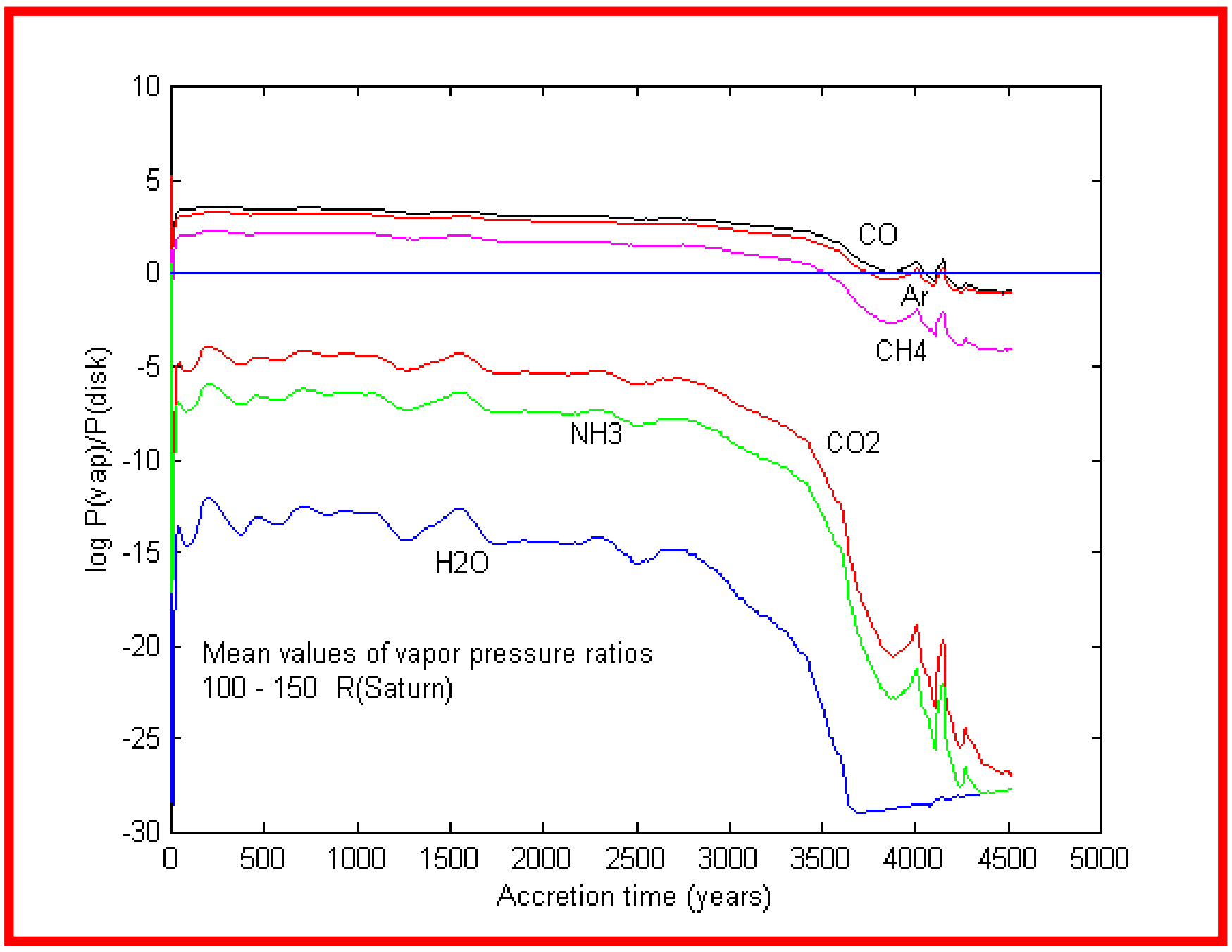

*Figure 3*

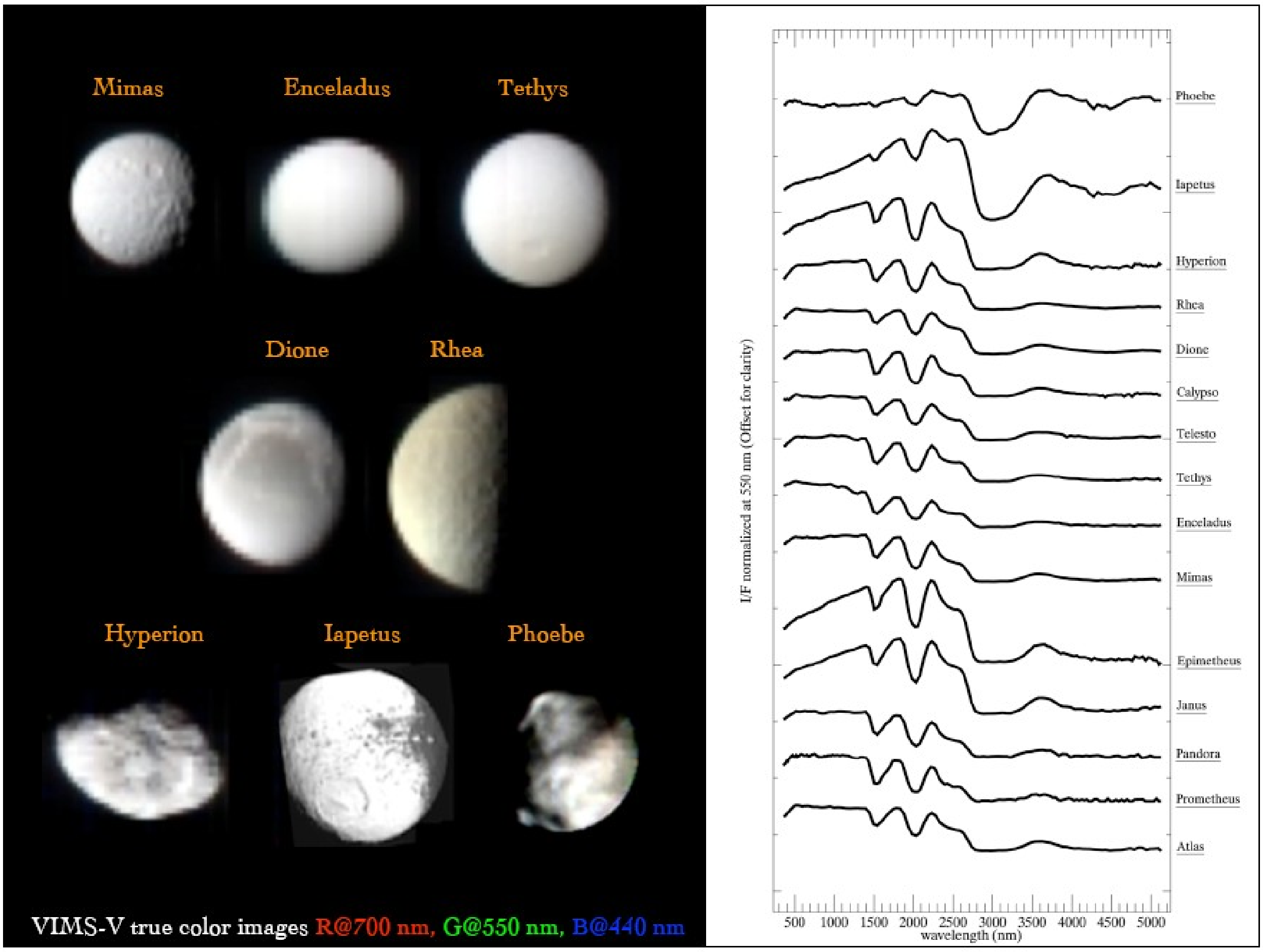

*Figure 4*

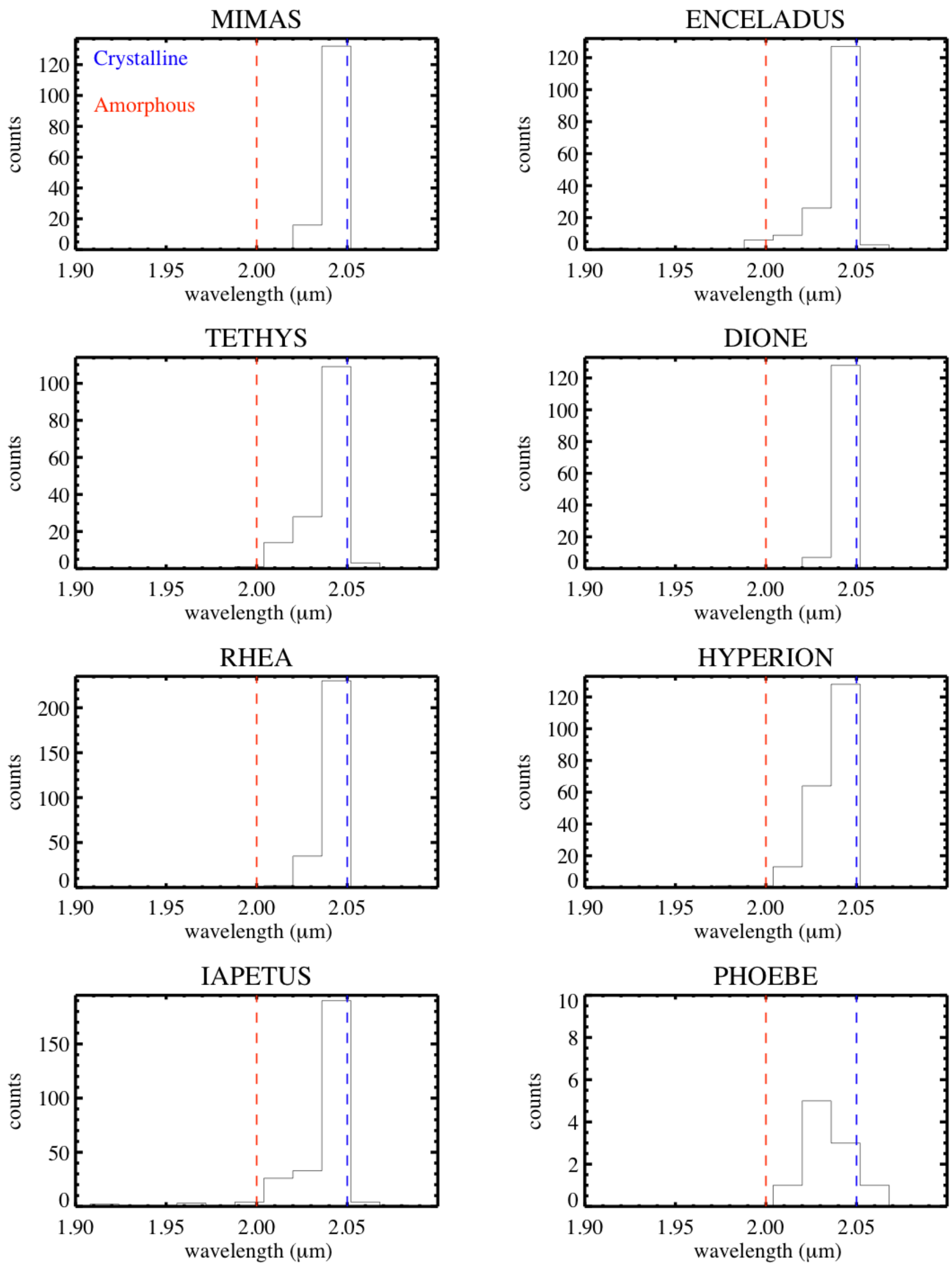
MIMAS
ENCELADUS
TETHYS
DIONE
RHEA
HYPERION
IAPETUS
PHOEBE
Crystalline
Amorphous
counts
wavelength (μm)
1.90
1.95
2.00
2.05

***Figure 5***

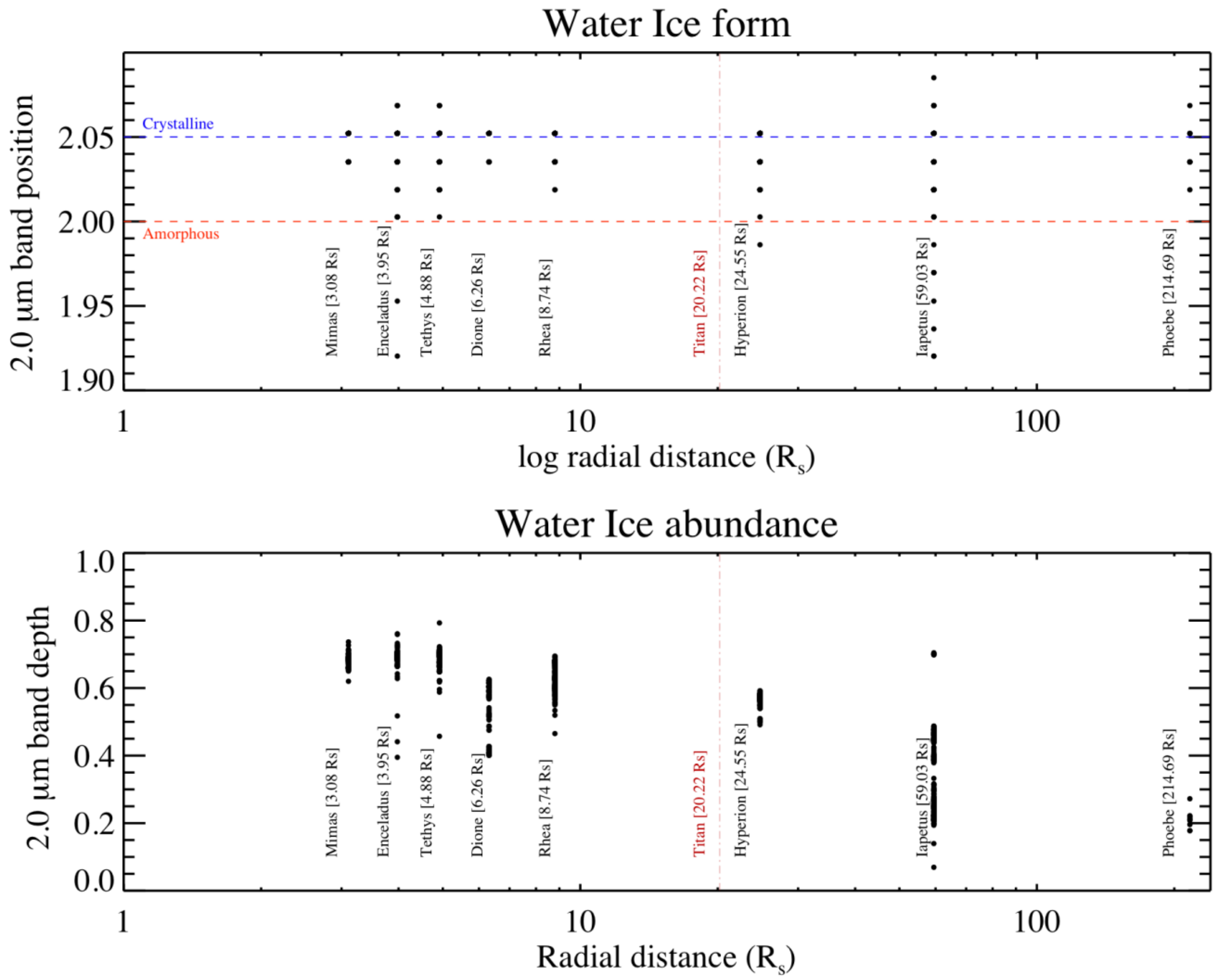

***Figure 6***

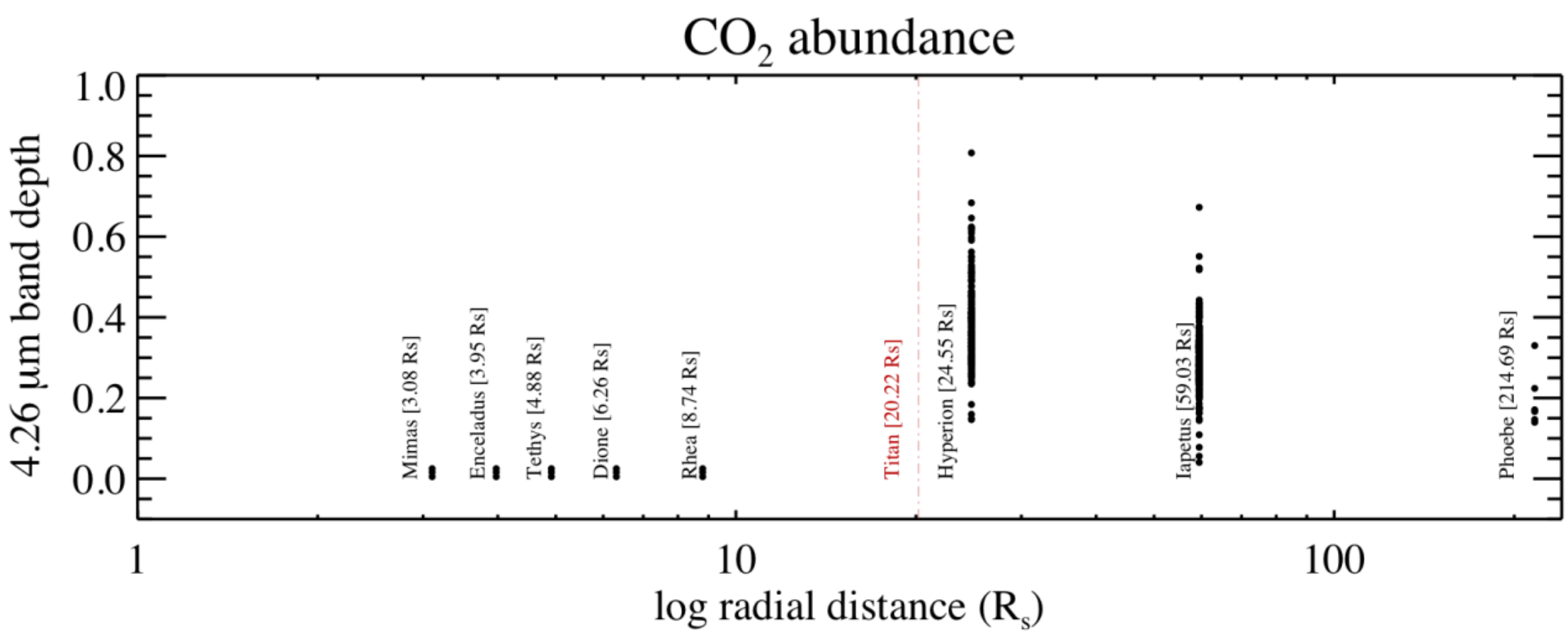

***Figure 7***

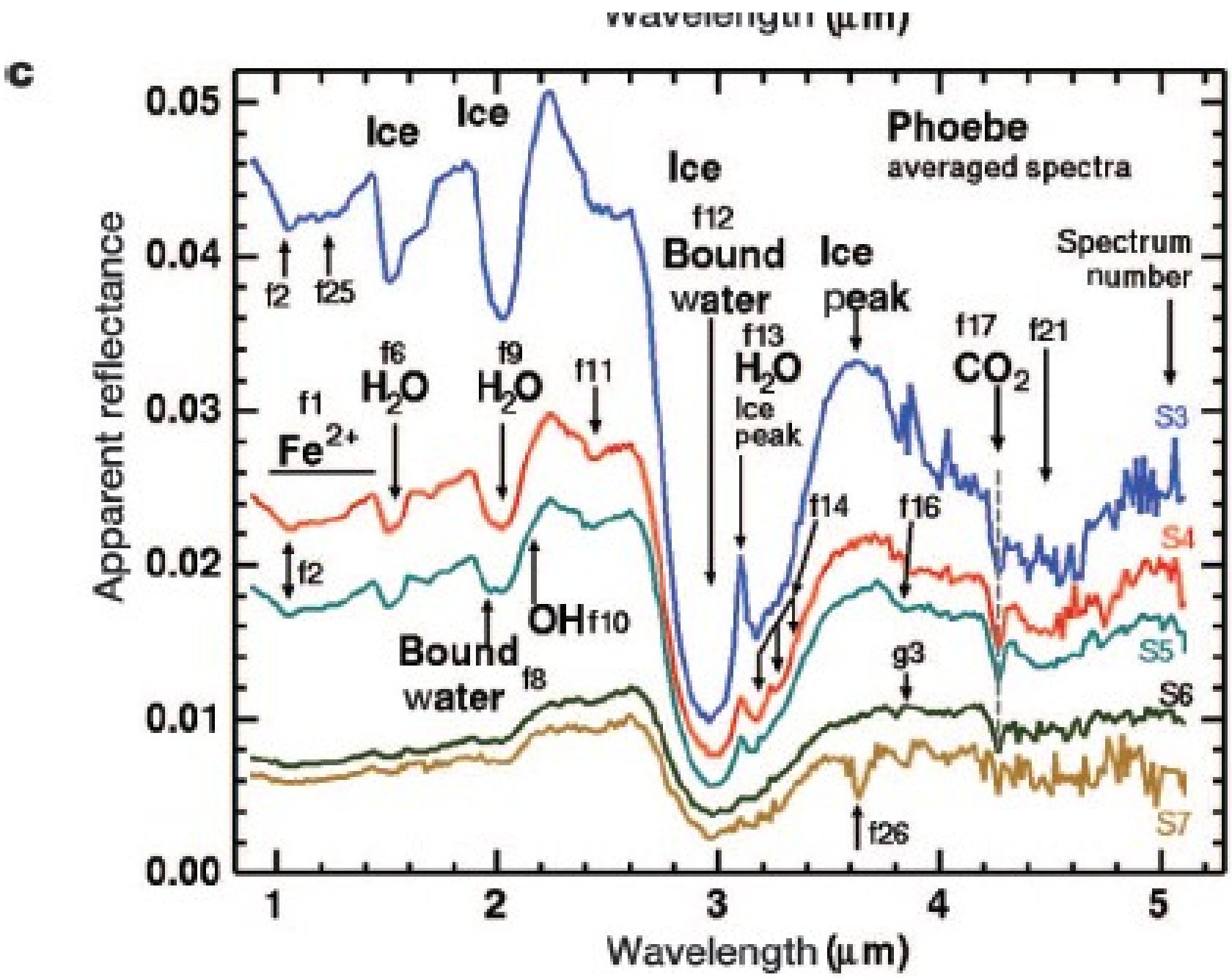

00

***Figure 8***

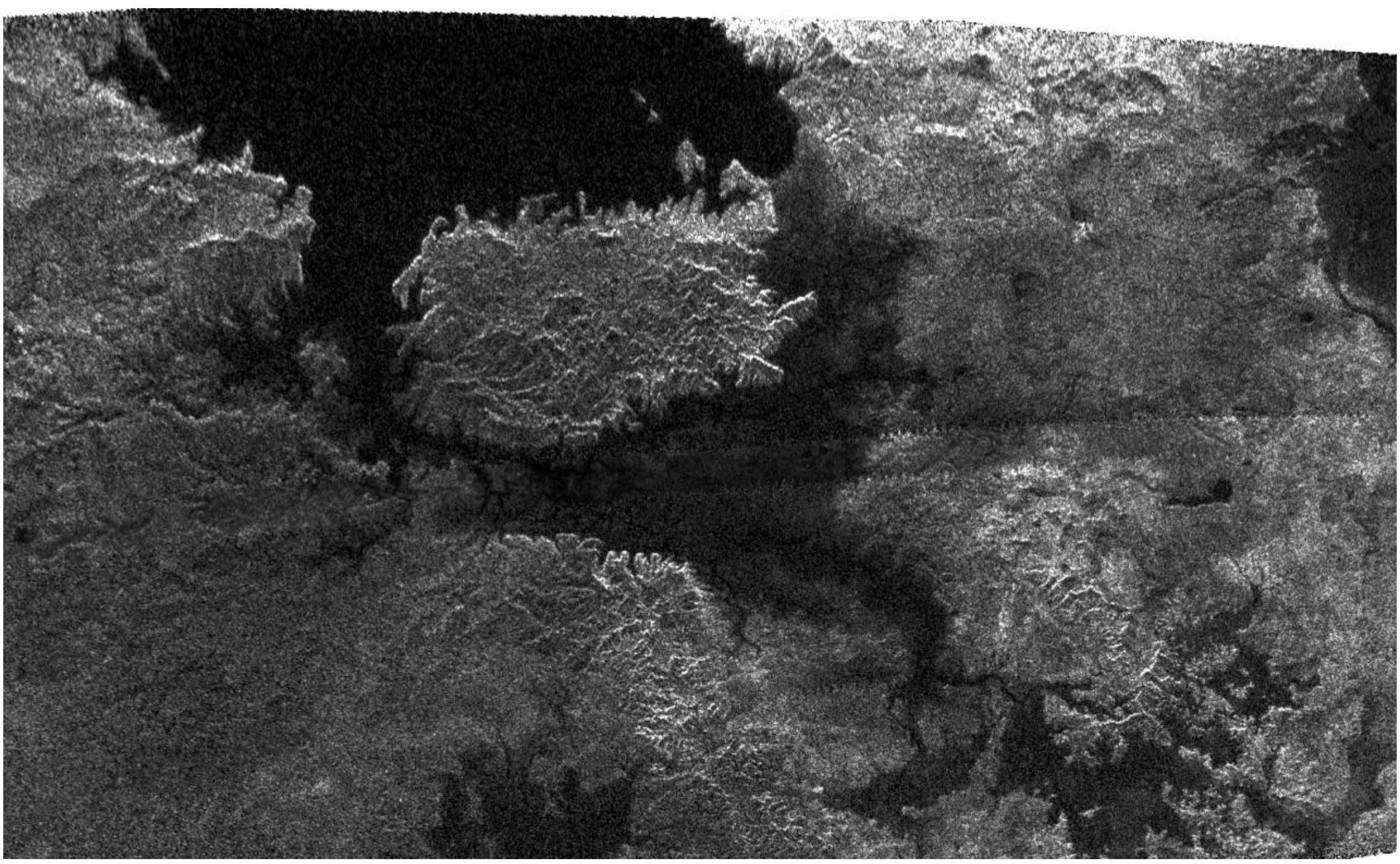

***Figure 9***

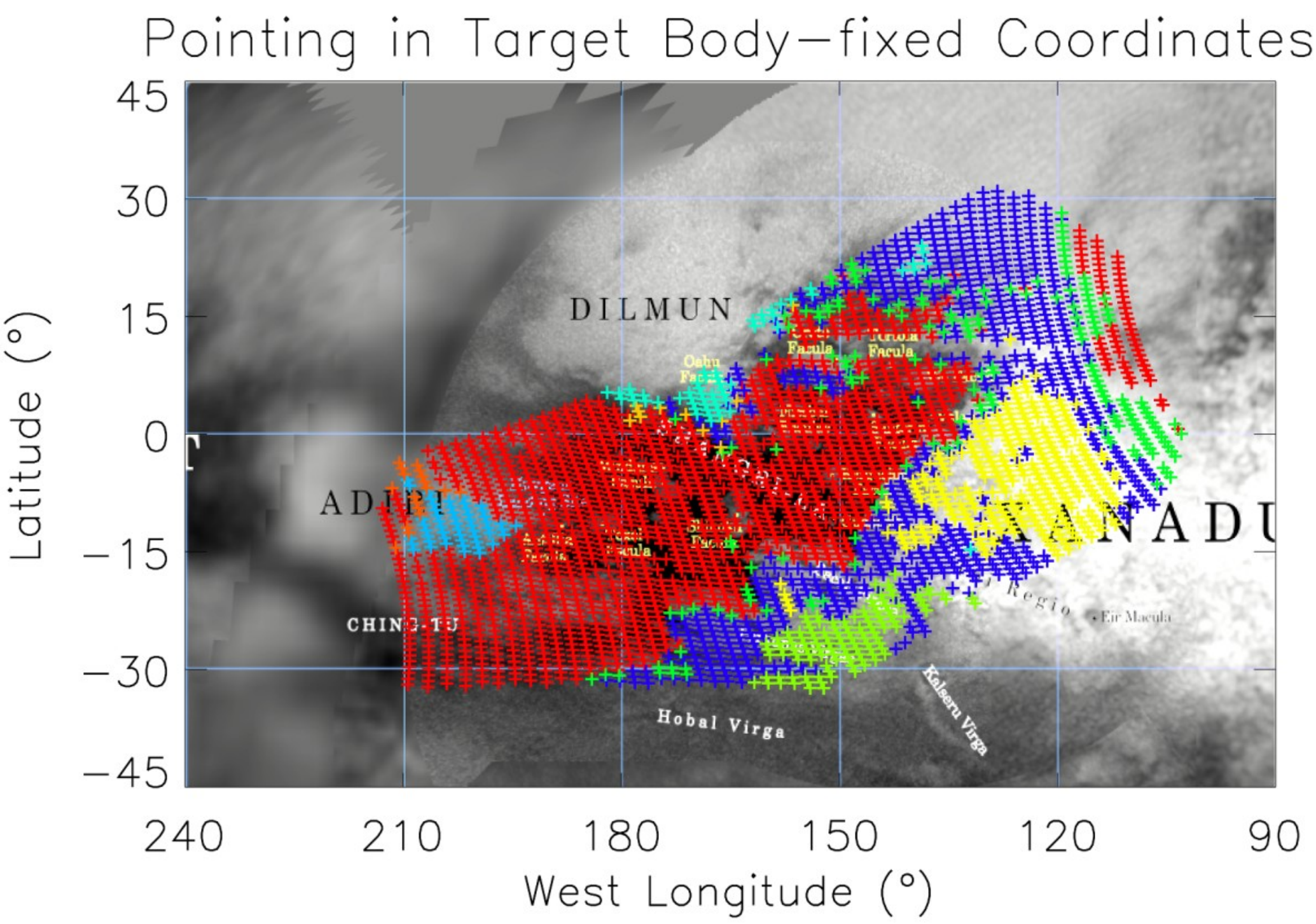

Pointing in Target Body−fixed Coordinates
Latitude (°)
West Longitude (°)
45
30
15
0
−15
−30
−45
240
210
180
150
120
90
DILMUN
ADIRI
Hobal Virga
Kaiseru Virga

***Figure 10***

## Figure caption

Figure 1 Satellite formation can be a natural byproduct of planet formation. In the figure the density distribution of the Saturn disk during the final phase of the accretion is shown. The density is expressed in logarithmic scale.

Figure 2 Disk phase diagram at different times. The initial phases are hot and dense, the disk is turbulent and the temperature rapidly decreases. When turbulence decreases, the final disk has low density and is cold (alpha parameter of the order of $10^{-4}$)

Figure 3 Average gas pressures in the Saturnian disk as a function of time. The pressure is computed as the average in the annular region between 100-150 Saturn radii. It is compared with the saturation pressure of different gases that can be present in the nebula. CO can be in form of ice only at the end of the accretion, while the other gases are always in condensed form.

Figure 4 Images of Saturn's satellites obtained by Cassini Camera (left) and average "full disk" spectra collected by Cassini VIMS.

Figure 5 Histograms of the H2O crystalline ice in the different satellites, compared to the amorphous ice. It is apparent that amorphous ice is only present in the satellites far from Saturn, or where "fresh ice" is generated.

Figure 6 Distribution of the crystalline and amorphous ice on the icy satellites based on the measure of the structure and strength of the water ice at 2□m.

Figure 7 Radial distributions of the "contaminants" across the Saturnian system measured based on the strength of 4.26 micrometers band. In the inner region the minimum is observed on the more uncontaminated object (Enceladus), reaching a maximum on Rhea and finally decreasing from Hyperion to Phoebe

Figure 8 Spectral diversity of Phoebe's surface. These spectra are averages from multiple small locations on Phoebe's surface, obtained from unprojected data.

Figure 9 This radar image, obtained by Cassini's radar instrument during a near-polar flyby on Feb. 22, 2007, shows a big island in the middle of one of the larger lakes imaged on Titan. The island is about 90 kilometers by 150 kilometers across. This image was taken in synthetic aperture mode at 700 meters resolution. North is toward the left-hand side. The image is centered at about 79 degrees north and 310 degrees west. Image Credit: NASA/JPL

Figure 10 From Tosi et al. (2009). Classification of VIMS cube CM1481607233_1 considering 5 variables: reflectance in the 2.02, 2.69, 2.78 and 5 μm windows, plus the normalized backscatter cross-section $\sigma 0$ as measured by RADAR. Confidence level: 93.64%. The results are superimposed to a map of Titan's surface derived by ISS data at 0.93 μm. Meaning of the colors: red = low reflectance in all windows and low backscattering coefficient; blue = medium reflectance in all windows and medium backscattering coefficient;

yellow = high reflectance and very high backscattering coefficient; green = high reflectance at 2.69, 2.78 and 5 μm and low backscattering coefficient; cyan = high reflectance at 2.02 μm and high backscattering coefficient.